\documentstyle[sprocl,epsfig]{article}

\def\be{\begin{equation}}
\def\ee{\end{equation}}
\def\bea{\begin{eqnarray}}
\def\eea{\end{eqnarray}}

\begin{document}

\title{RANDOMLY INTERACTING BOSONS, MEAN-FIELDS AND $L=0$ GROUND STATES}

\author{R. BIJKER$^1$, A. FRANK$^{1,2}$}

\address{$^1$ICN-UNAM, AP 70-543, 04510 M\'exico, DF, M\'exico\\
$^2$CCF-UNAM, AP 139-B, Cuernavaca, Morelos, M\'exico}

\maketitle

\abstracts{Random interactions are used to investigate to what extent 
the low-lying behavior of even-even nuclei depend on particular 
nucleon-nucleon interactions. The surprising results that were 
obtained for the interacting boson model, i.e. the dominance of 
ground states with $L=0$ and the occurrence of both vibrational and 
rotational structure, are interpreted and explained in terms of a 
mean-field analysis.}

\section{Introduction}

In empirical studies of medium and heavy even-even nuclei very regular 
features have been observed, such as the tripartite classification of 
nuclear structure into seniority, anharmonic vibrator and rotor regions 
\cite{Zamfir}. In each of these three regimes, the energy systematics is 
extremely robust, and the transitions between different regions occur very 
rapidly, typically with the addition or removal of only one or two pairs 
of nucleons. Traditionally, this regular behavior has been interpreted 
as a consequence of particular nucleon-nucleon interactions, such as an
attractive pairing force in semimagic nuclei and an attractive
neutron-proton quadrupole-quadrupole interaction for deformed nuclei. 

It came as a surprise, therefore, that recent shell model studies of 
even-even nuclei with two-body random interactions (TBRE) displayed a 
marked statistical preference for ground states with $L=0$, energy 
gaps, and other signals of ordered behavior \cite{JBD}. This work 
has sparked a large number of investigations, both in fermion 
\cite{BFP1,JBDT,BFP2,MVZ,ZA} and in boson systems 
\cite{BFP2,BF1,BF2,KZC,DK,BF3}, in order to further explore and to 
explain these remarkable and unexpected results. 

These robust features suggest that there exists an underlying simplicity 
of low-energy nuclear structure never before appreciated. In this 
contribution, we discuss the results of a study of the systematics 
of collective levels in the framework of the interacting boson model (IBM) 
with random interactions \cite{BF1,BF2}, and present a possible explanation 
in terms of a mean-field analysis. 

\section{Randomly interacting bosons}

\begin{figure}
\centerline{\hbox{\epsfig{figure=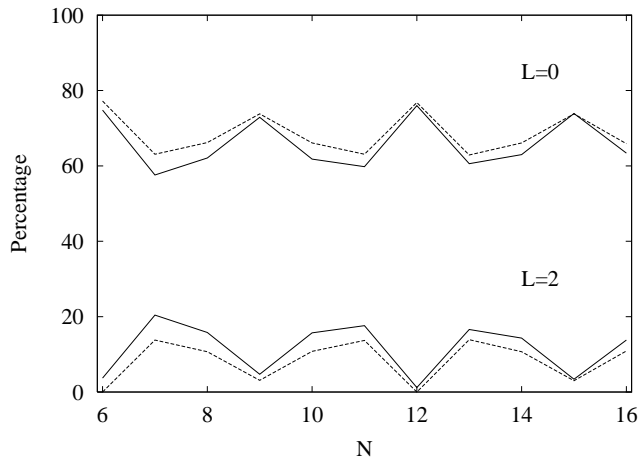,width=0.75\linewidth} }}
\caption[]{Percentages of ground states with $L=0$ and $L=2$ in the IBM 
with random one- and two-body interactions calculated exactly for 
1,000 runs (solid lines) and in mean-field approximation (dashed lines).}
\label{ibmgs}
\end{figure}

In the IBM, collective excitations in nuclei are described in 
terms of a system of $N$ interacting monopole and quadrupole 
bosons \cite{IBM}. We consider the most general one- and two-body 
Hamiltonian. The two one-body and seven 
two-body matrix elements are chosen independently from 
a Gaussian distribution of random numbers with zero mean and width 
$\sigma$, such that the ensemble is invariant under orthogonal basis 
transformations \cite{BF1,BF2}. 
For each set of randomly generated one- and two-body matrix elements 
we calculate the entire energy spectrum and analyze the results. 

\begin{figure}
\centerline{\hbox{\epsfig{figure=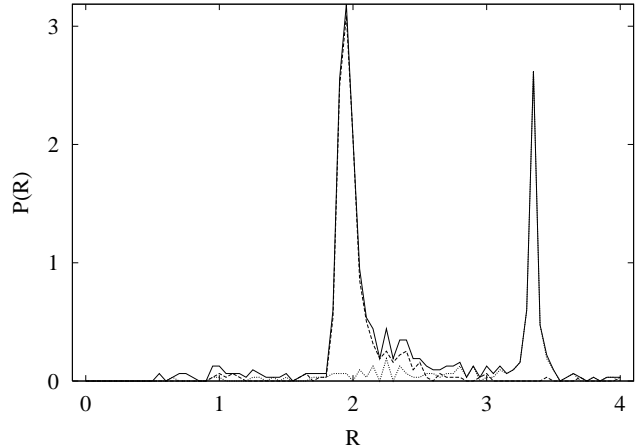,width=0.75\linewidth} }}
\caption[]{Probability distribution $P(R)$ of the energy ratio of 
Eq.~(\protect\ref{rat}) obtained for $N=16$ and 1,000 runs (solid line). 
The dashed and dotted lines show the contribution of the spherical 
and deformed equilibrium shapes, respectively.}
\label{ratio}
\end{figure}

In Fig.~\ref{ibmgs} we show the percentages of $L=0$ and $L=2$ ground 
states as a function of the total number of bosons $N$ (solid lines). 
We see a clear dominance of ground states with $L=0$ with 
$\sim$ 60-75 $\%$. Both for $L=0$ and $L=2$ there are large oscillations 
with $N$. For $N=3n$ (a multiple of 3) we see an enhancement for $L=0$ 
and a decrease for $L=2$. The sum of the two hardly depends on the 
number of bosons. 

For the cases with a $L=0$ ground state, we show in Fig.~\ref{ratio} 
the probability distribution $P(R)$ of the energy ratio for $N=16$ 
(solid line) 
\bea
R &=& [E_{4_1}-E_{0_1}]/[E_{2_1}-E_{0_1}] ~. 
\label{rat}
\eea
There are two very pronounced 
peaks, right at the vibrational value of $R=2$ and at the rotational 
value of $R=10/3$, a clear indication of the occurrence of vibrational 
and rotational structure. This has been confirmed by a simultaneous 
study of the quadrupole transitions between the levels \cite{BF1}. 

These are surprising results in the sense that, according to the 
conventional ideas in the field, the occurrence of $L=0$ ground states 
and the existence of vibrational and rotational bands are due to very 
specific forms of the interactions. The study of the IBM with random 
interactions seems to indicate that this may not be the entire story. 
However, the above results were obtained from numerical studies. 
It would be very interesting to gain a better understanding as to  
why this happens. What is the origin of the regular features which  
arise from random interactions? 
In this respect, there is a relevant quote by E.P. Wigner (as 
communicated to us by M. Moshinsky): {\it `I am happy to learn that the 
computer understands the problem, but I would like to understand it too'}.  

A first attempt in this direction was made for the vibron model, 
which has many of the same qualitative features as the IBM, 
but has a much simpler mathematical 
structure. We showed that the emergence of regular features from the 
vibron model with random interactions is related to the existence of 
three different geometric shapes \cite{BF4}. In the next section, we 
carry out a similar mean-field study of the IBM. 

\section{Mean-field analysis}

The connection between the IBM, potential energy surfaces, equilibrium 
configurations and geometric shapes, can be studied by means of 
coherent states. The coherent state for the IBM can be written 
as a condensate of a deformed boson which is a superposition of a scalar 
and a quadrupole boson \cite{cs1,cs2}
\bea
\left| \, N,\alpha \, \right> \;=\; \frac{1}{\sqrt{N!}} \, 
\left( \sqrt{1-\alpha^2} \, s^{\dagger} + \alpha \, d_0^{\dagger} 
\right)^N \, \left| \, 0 \, \right> ~, 
\label{trial}
\eea
with $-1 < \alpha \leq 1$. The potential energy surface is then given 
by the expectation value of the Hamiltonian in the coherent state 
\bea
E_N(\alpha) \;=\; \left< \, N,\alpha \, \right| \, H \, \left| \, 
N,\alpha \, \right> \;=\; a_4 \, \alpha^4 + 
a_3 \, \alpha^3 \sqrt{1-\alpha^2} + a_2 \, \alpha^2 + a_0 ~, 
\label{surface}
\eea
where the coefficients $a_i$ are linear combinations of the parameters 
of the Hamiltonian which in turn are taken as independent random numbers. 

For random interactions, we expect the trial wave function of 
Eq.~(\ref{trial}) and the energy surface of Eq.~(\ref{surface}) to 
provide information on the distribution of shapes that the model can 
acquire. The equilibrium configuration is characterized by the value of 
$\alpha=\alpha_0$ for which the energy surface $E_N(\alpha)$ has its 
minimum value. For a given Hamiltonian, the value of $\alpha_0$ depends 
on the coefficients $a_4$, $a_3$ and $a_2$. The distribution of shapes for 
an ensemble of Hamiltonians then depends on the joint probability 
distribution $P(a_4,a_3,a_2)$. 
In this approximation, we find that there are only three possible 
equilibrium configurations: 
\begin{itemize}

\item $\alpha_0=0$: $s$-boson condensate. This corresponds to a spherical 
shape which can only have $L=0$. 

\item $0 < \alpha_0 < 1$ or $-1 < \alpha_0 < 0$: deformed condensate with 
prolate or oblate symmetry, respectively. A deformed shape corresponds to 
a rotational band with angular momenta $L=0,2,\ldots,2N$. The ordering 
of the energy levels is determined by the sign of the moment of inertia 
\bea
E_{\rm rot} &=& \frac{1}{2{\cal I}_3} L(L+1) ~. 
\eea

\item $\alpha_0=1$: $d$-boson condensate. 
The rotational structure of a $d$-boson condensate is more complicated. 
It is characterized by the labels $\tau$, $n_{\Delta}$ and $L$. The boson 
seniority $\tau$ is given by $\tau=3n_{\Delta}+\lambda=N,N-2,\ldots,1$ or 
$0$ for $N$ odd or even, and the values of the angular momenta are 
$L=\lambda,\lambda+1,\ldots,2\lambda-2,2\lambda$ \cite{IBM}. In this case, 
the rotational excitation energies depend on two moments of inertia 
\bea
E_{\rm rot} &=& \frac{1}{2{\cal I}_5} \tau(\tau+3) 
+ \frac{1}{2{\cal I}_3} L(L+1) ~, 
\eea
which are associated with the spontaneously broken three- and 
five-dimensional rotational symmetries of the $d$-boson condensate. 

\end{itemize}
The probability that the ground state of each of these equilibrium 
configurations has $L=0$ can be estimated by evaluating simultaneously 
the corresponding moments of inertia (e.g. with the Thouless-Valatin 
prescription). 

In Fig.~\ref{ibmgs} we show the percentages of $L=0$ and $L=2$ ground 
states, as calculated in the mean-field analysis (dashed lines). 
A comparison with the exact results (solid lines) shows an 
excellent agreement. The spherical and deformed condensates contribute 
constant amounts of 39.4 $\%$ and 23.4 $\%$, respectively, to the $L=0$ 
ground state percentage. The oscillations observed for $L=0$ are 
therefore entirely 
due to the rotational structure of the $d$-boson condensate. The $L=2$ 
ground states arise completely from the $d$-boson condensate solution. 
In the mean-field analysis the ground state has $L=0$ or $L=2$ in 
$\sim$ 77 $\%$ of the cases. For the remaining 23 $\%$ of the cases the 
ground state has the maximum value of the angular momentum $L=2N$. 
This percentage is almost a constant and hardly depends on $N$, in 
agreement with the exact results. 

In Fig.~\ref{ratio} we show the contribution of each of the equilibrium 
configurations to the probability distribution of the energy ratio 
of Eq.~(\ref{rat}). We see that the spherical shape (dashed line) 
contributes almost exclusively to the peak at $R=2$, and similarly the 
deformed shape (dotted line) to the peak at $R=10/3$, which once again 
confirms the vibrational and rotational character of these maxima. 
For $N=16$ the contribution of the $d$-boson condensate is small. 

\section{Summary and conclusions}

In this contribution, we have studied the properties of low-lying
collective levels in the IBM with random interactions. We addressed 
the origin of the regular features, that had been obtained before in 
numerical studies, in particular the dominance of $L=0$ ground states 
and the occurrence of vibrational and rotational band structures. 

It was shown that a mean-field analysis of the IBM with random 
interactions can account for all of these features. They are related to 
the existence of three different equilibrium configurations or geometric 
shapes: a spherical shape ($\sim$ 39 $\%$), a deformed shape 
($\sim$ 36 $\%$) and a condensate of quadrupole bosons ($\sim$ 25 $\%$). 
Since the spherical shape only has $L=0$, and the deformed shape in 
about two thirds of the cases, these two solutions account for 
$\sim$ 63 $\%$ of $L=0$ ground states. The oscillations observed 
for the $L=0$ ground state percentage can be ascribed totally to 
the contribution of the $d$-boson condensate. 
Finally, we found a one-to-one correspondence between the peaks in the 
probability distribution for the energy ratio and the occurrence of 
the spherical and deformed equilibrium configurations.  

In summary, the use of mean-field techniques allows one to associate 
different regions of the parameter space with geometric shapes. 
This method bypasses the diagonalization of thousands of matrices, 
and provides an explanation of all regular features that have been 
observed in studies of the IBM with random interactions. 

\section*{Acknowledgements}

It is a great pleasure to dedicate this contribution to the 
65th birthday of Peter von Brentano. Happy birthday! 
This work was supported in part by CONACyT under projects 
32416-E and 32397-E, and by DPAGA under project IN106400.

\end{document}